\documentclass[12pt]{article}
\usepackage{amsmath,amsfonts,amssymb,amsthm,bm}
\usepackage{natbib}
\usepackage{graphicx}
\usepackage{color}
\usepackage{subfigure}

\usepackage{setspace}
\theoremstyle{definition}
\newtheorem{defn}{Definition}

\theoremstyle{remark}

\theoremstyle{plain}
\newtheorem{propn}[defn]{Proposition}

\newcommand{\kth}[1]{${#1}^{\textrm{th}}$}

\begin{document}
\singlespacing
\title{Proximity penalty priors for Bayesian mixture models} 
\author{Matthew Sperrin\\[4.1pt]
{\it \small Department of Mathematics and Statistics, Lancaster University, UK}\\[2.1pt]
{\it \small m.sperrin@lancaster.ac.uk}
%\author{Blind}
\\
}

\maketitle

\begin{abstract}
When using mixture models it may be the case that the modeller has a-priori beliefs or desires about what the components of the mixture should represent. For example, if a mixture of normal densities is to be fitted to some data, it may be desirable for components to focus on capturing differences in location rather than scale. We introduce a framework called proximity penalty priors (PPPs) that allows this preference to be made explicit in the prior information. The approach is scale-free and imposes minimal restrictions on the posterior; in particular no arbitrary thresholds need to be set. We show the theoretical validity of the approach, and demonstrate the effects of using PPPs on posterior distributions with simulated and real data.


\end{abstract}

\section{Introduction}
Mixture models are widely recognized as a useful tool for inference in a variety of settings.
Having been first used over 100 years ago \citep[for example, in][]{pearson1894}, more recently mixture models are enjoying a revival, thanks to advances in computational methods for inference. In particular, the EM algorithm \citep{dempster77} and MCMC \citep[see, for example,][]{diebolt94} have driven considerable advances in the field.
See \cite{mclachlan00} for a general overview of mixture models; \cite{schnatter06} provides an overview of Bayesian mixture models, which are the focus of this paper.

We recall the definition of a mixture model and introduce notation.
Suppose $n$ observations, $y_1,\ldots,y_n$, are taken from a $K$-component mixture distribution where all the components have the same distributional form, with mixture-specific parameters $\bm{\theta}=(\bm{\theta}_1,\ldots,\bm{\theta}_K)$, global parameters $\bm{\eta}$ and mixing weights $\bm{\pi}=(\pi_1,\ldots,\pi_K)$, summarised by $\bm{\gamma} = (\bm{\pi}, \bm{\theta}, \bm{\eta})$.
The mixture distribution for a single observation $Y_i$ is then given by
\begin{equation}\label{eq:firstform}
g\left(y_i|\mathbf{\bm{\gamma}}\right) = \sum_{k=1}^K \pi_k f_k\left(y_i|\bm{\theta}_k,\bm{\eta}\right),
\end{equation}
with $K \geq 1$, $\pi_k > 0 \  (k=1,2,\ldots,K)$, $\sum_{k=1}^K\pi_k = 1$
and $f_k(\cdot|\bm{\theta}_k,\bm{\eta})$ is a density function parametrised by $\bm{\theta}_k$ and $\bm{\eta}$.

A Bayesian approach to estimating the parameters of the mixture distribution of Equation (\ref{eq:firstform}) involves the specification of priors for the parameters $\bm{\gamma}$. The issue of prior specification in this context has a number of difficulties. 

First, fully improper priors cannot be used for component-specific parameters in mixture models, since doing so causes the posterior to be improper also  \citep[see, for example,][]{mclachlan00}. However, proper priors, even with large variance, can have considerable influence on the posterior distribution, and the extent of this influence can be difficult to assess \citep{marin05}.
Re-parametrisation in a hierarchical manner and allowing only the global parameters to be improper is one solution: this is considered by \cite{mengersen96}, and \cite{roeder97}.  Another possibility is to use data-dependent priors, as considered by \cite{richardson97}, and \cite{wasserman00}. 

Second, where no component specific information is available, identical priors may be proposed for the components of each parameter. This leads to a non-identifiable posterior, which is known as the label switching problem. This has been well studied \citep[see, for example][and references therein]{stephens00,jasra05,sperrin09labels}.

Third, constructing independent priors for component parameters may not be sensible, as the components only have meaning relative to one another \citep{lee08}.

This third issue is the focus of this paper. We consider in detail the idea that priors should be specified relative to each other. We introduce a strategy for doing so that we call `proximity penalty priors' (PPPs). The basic idea is that priors are specified in two parts: first, each prior is specified independently, corresponding to standard existing approaches; second, a proximity penalty is applied, which penalises the joint prior distribution of certain configurations of parameters. We show that the construction makes theoretical sense.

Section \ref{sec:ppp} introduces the idea of PPPs. Section \ref{sec:results} illustrates the consequences of the PPP approach on real and simulated data; the paper concludes with a discussion in Section \ref{sec:discuss}.

\section{Proximity Penalty Priors} \label{sec:ppp}

We begin with a simple result that establishes the validity of the PPP approach.
\begin{propn} \label{ppthm}
Suppose the prior for $\bm{\gamma}$, given by $p(\bm{\gamma})$,
can be separated as
\begin{equation*}
p(\bm{\gamma}) = p_1(\bm{\gamma}) p_2(\bm{\gamma}).
\end{equation*}
Denote the likelihood by $L(\bm{\gamma})$ and the posterior by $q(\bm{\gamma})$, so that $q(\bm{\gamma}) \propto L(\bm{\gamma})p(\bm{\gamma})$.
Suppose that a new parameter vector $\bm{\gamma}^*$ can be simulated from a proposal distribution $r(\bm{\gamma^*}) = L(\bm{\gamma^*}) p_1(\bm{\gamma^*})$, and the existing value of $\bm{\gamma}$ is 
$\bm{\gamma}^{m}$.
Then if we set
\begin{equation}
\bm{\gamma}^{m+1} = \left\{ \begin{array}{ll}
\bm{\gamma}^* &\textrm{with probability} \ \min\left(1,\frac{p_2(\bm{\gamma}^*)}{p_2(\bm{\gamma}^{m})}\right)\\
\bm{\gamma}^{m} &\textrm{otherwise},
\end{array}\right.
\end{equation}
\\
the result is equivalent to a Metropolis-Hastings update. 

\end{propn}

\proof The acceptance probability for the Metropolis-Hastings procedure with proposal density $r(\cdot)$ and posterior $q(\cdot)$ is
\begin{equation*}
\min\left(1,\frac{q(\bm{\gamma}^*) r(\bm{\gamma}^{m})}{q(\bm{\gamma}^{m}) r(\bm{\gamma}^*)} \right).
\end{equation*}
Substituting in these densities gives the result.
\qed

In the context of this work the portion of the prior $p_1(\cdot)$ corresponds to the independent specification of the parameters, for which standard distributions could be used; the portion $p_2(\cdot)$ corresponds to the novel part of the prior that jointly assesses the values of the parameters and penalises undesirable combinations.

Suppose that the priors $p_1(\cdot)$ are conjugate. Then an MCMC approach would proceed, on each iteration, by generating proposed new parameters according to a Gibbs sampling scheme with the full conditionals based on the prior component $p_1(\cdot)$, then accepting the proposed parameters according to a Metropolis Hastings ratio on the prior component $p_2(\cdot)$.

We illustrate the idea with an example.
Consider a mixture of two normal distributions
\begin{equation} \label{eq:normmix}
p\left(y_i|\mathbf{\bm{\gamma}}\right) = \pi_1 N(y_i; \mu_1,\sigma_1^2) + \pi_2 N (y_i; \mu_2, \sigma_2^2),
\end{equation}
with $\pi_1 + \pi_2 = 1$, and all the parameters $\bm{\gamma} = (\pi_1,\pi_2,\mu_1,\mu_2,\sigma_1^2,\sigma_2^2)$ unknown. Standard conjugate prior choices would then be a Dirichlet distribution for the pair $(\pi_1,\pi_2)$, normal distributions for $\mu_1$ and $\mu_2$, and inverse-gamma distributions for $\sigma_1^2$ and $\sigma_2^2$. Throughout this paper we will use the empirical Bayes prior distributions suggested by \cite{richardson97} unless otherwise stated.
We may believe a-priori that the key difference between the two components is the location. If the components are not well separated or the amount of data is small it is important that such prior information is captured. By Proposition 1, we can reflect these beliefs in a separate part of the prior $p_2(\cdot)$. A sensible such choice is
\begin{equation} \label{eq:simple-mu}
p_2(\bm{\gamma}) = |\mu_1 - \mu_2|.
\end{equation}
Such a function assigns more prior weight to larger differences between $\mu_1$ and $\mu_2$.
In isolation, the above $p_2(\cdot)$ is improper but provided $p_1(\cdot)$ is proper the overall prior is proper.
Such a prior enjoys scale invariance in the sense that $p_2(a\bm{x}_1)/p_2(a\bm{x}_2) = p_2(\bm{x}_1)/p_2(\bm{x}_2)$ for all non-zero $a$. This may or may not be desirable. An alternative would be to specify a distance $\delta$ as a minimum distance between $\mu_1$ and $\mu_2$, i.e.
$$
p_2(\bm{\gamma}) = \bm{1}_{(|\mu_1 - \mu_2| > \delta)}.
$$
This generates the question of how $\delta$ should be specified, but may be appropriate in some situations.

More generally, for a mixture distribution with $K$ parameters, suppose there exists a component-specific parameter $\phi_k$ for each component $k=1,\ldots,K$, and the difference between the components is a-priori believed (or, from the point of view of model interpretation, desired) to be in terms of this parameter. Then we propose setting
\begin{equation} \label{eq:ppp-diff}
p_2(\bm{\gamma}) = \min_{k \neq l} |\phi_k - \phi_l|.
\end{equation} 
On the other hand, for a mixture distribution with $K$ parameters, if there exists a component-specific parameter $\psi_k$ for each component $k=1,\ldots,K$, and each component is a-priori expected or desired to have \emph{similar} values of this parameter, we could set
\begin{equation} \label{eq:ppp-same}
{p}_2(\bm{\gamma}) = \max_{k \neq l} |\psi_k - \psi_l|^{-1}.
\end{equation}
Here, the scale free nature of $p_2(\cdot)$ is an advantage in that we do not have to quantify `similar'.
More generally, $p_2(\bm{\gamma})$ could be constructed as any multiplicative combination of Equations (\ref{eq:ppp-diff}) and (\ref{eq:ppp-same}).
The procedure can also be applied when the number of components $K$ is allowed to vary, in which case it makes sense only within fixed values of $K$ in the same way that the label switching problem only has meaning within fixed values of $K$ \citep{nobile07}.

\section{Examples} \label{sec:results}
\subsection{Mixture of Two Normals}
Our first illustration takes the simple mixture of two normals example. We generate 100 observations from the density given in Equation (\ref{eq:normmix}), with $\mu_1 =0$,
$\mu_2 = 2$, $\sigma_1^2 = \sigma_2^2 = 1$ and $\pi_1 = \pi_2 = 0.5$. We consider two prior specifications: 
\begin{enumerate}
\item[(a)] the standard specification given in \cite{richardson97}, denoted \emph{without PPP}; 
\item[(b)] a two part prior $p(\bm{\gamma}) = p_1(\bm{\gamma}) p_2(\bm{\gamma})$, with $p_1(\bm{\gamma})$ as given in \cite{richardson97} and $p_2(\bm{\gamma})$ as given in Equation (\ref{eq:simple-mu}), denoted \emph{with PPP}.
\end{enumerate}
In both cases we fix the number of components $K=2$. In (b), we are therefore adding an explicit prior opinion that the difference between the two components is in the locations $\mu_1$ and $\mu_2$.

\begin{figure}[htp]
\centering

\subfigure[without PPP]{
   \includegraphics[scale=0.5] {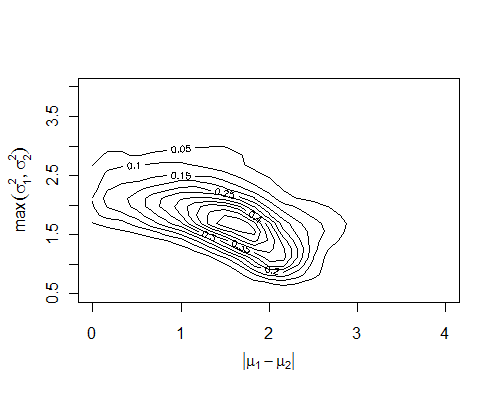}
   %\label{fig:maxsig-np}
 }

 \subfigure[with PPP]{
   \includegraphics[scale=0.5] {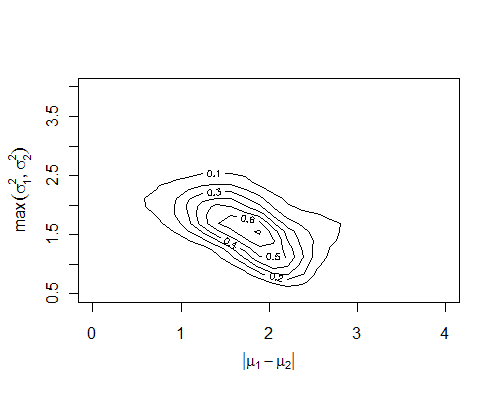}
   %\label{fig:maxsig-p}
 }

\caption{Posterior contour plots of $|\mu_1 - \mu_2|$ versus max$(\sigma_1^2,\sigma_2^2)$}
\label{fig:maxsig}
\end{figure}

\begin{figure}[htp]
\centering

    \includegraphics[scale=0.5] {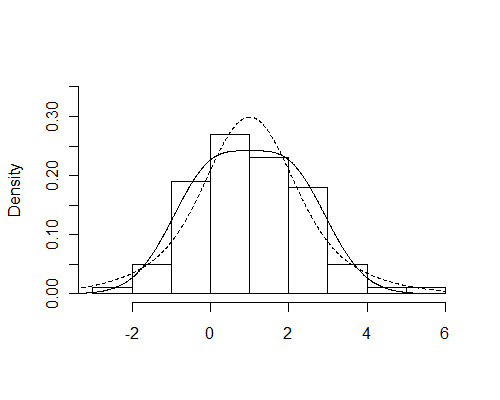}

\caption{Histogram of 100 realisations from $0.5N(0,1) + 0.5N(2,1)$ with true density overlaid (solid line) and alternative density, $0.5N(1,1) + 0.5N(1,4)$ also overlaid (dashed line)}
\label{fig:twodensities}
\end{figure}

Figure \ref{fig:maxsig} compares a bivariate projection of the posterior onto the absolute difference $|\mu_1 - \mu_2|$ and max$(\sigma_1^2,\sigma_2^2)$  without and with the PPP. Without the PPP, posterior mass is assigned to the situation where $|\mu_1 - \mu_2|$ is small and max$(\sigma_1^2,\sigma_2^2)$ is large. This corresponds to a case where a mixture distribution with similar means but different variances is fitted. In Figure \ref{fig:twodensities} we see that such a mixture is well supported by the data (dashed line in the figure). Once the PPP is applied, far less posterior mass is assigned to this scenario, since our prior distribution specifically tells us to exclude such cases.

%%%%%%%%%%%%%%%%%%%%%%%%%%%%%%%

%%%%%%%%%%%%%%%%%%%%%%%%%%%%%%%

%%%%%%%%%%%%%%%%%%%%%%%%%%%%%%%

\begin{figure}[htp]
\centering
\subfigure[without PPP]{
   \includegraphics[scale=0.5] {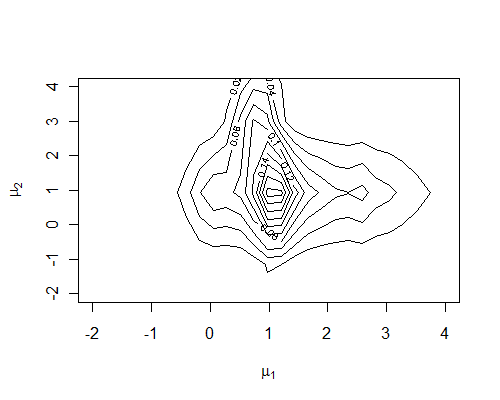}
   %\label{fig:mu1mu2-np}
 }

 \subfigure[with PPP]{
   \includegraphics[scale=0.5] {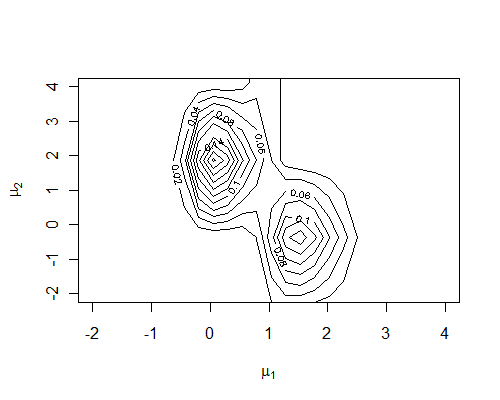}
   %\label{fig:mu1mu2-p}
 }

\caption{Posterior contour plots of $\mu_1$ versus $\mu_2$}
\label{fig:mu1mu2}
\end{figure}

Figure \ref{fig:mu1mu2} gives the marginal bivariate posterior of $(\mu_1,\mu_2)$, with and without the PPP. Without the PPP, the posterior appears to have a single mode at approximately $\mu_1 = \mu_2 = 1$; with the PPP, the posterior is bimodal with modes at approximately ($\mu_1 = 0, \mu_2=2$) and ($\mu_1=2, \mu_2 = 0$). The bimodality in the PPP case is a consequence of label switching; if component-specific inference is required, post-hoc relabelling should be carried out \citep[see, for example,][]{sperrin09labels}. The unimodality in the non PPP case is caused by the two means being very close together and the variances to differ, corresponding to a different interpretation of the mixture components.

%%%%%%%%%%%%%%%%%%%%%%%%%%%%%%%

We also ran the same comparison without assuming a fixed number of components $K$ \citep[using the birth-death method of][]{stephens00}, putting a Poisson$(1)$ prior distribution on the number of components $K$ \citep[see][for a justification of the use of this prior]{nobile07}. Similar results to the above were observed when we looked at the output conditional on $K=2$.

\subsection{Galaxy Data}
\begin{figure}[htp]
\centering

    \includegraphics[scale=0.5] {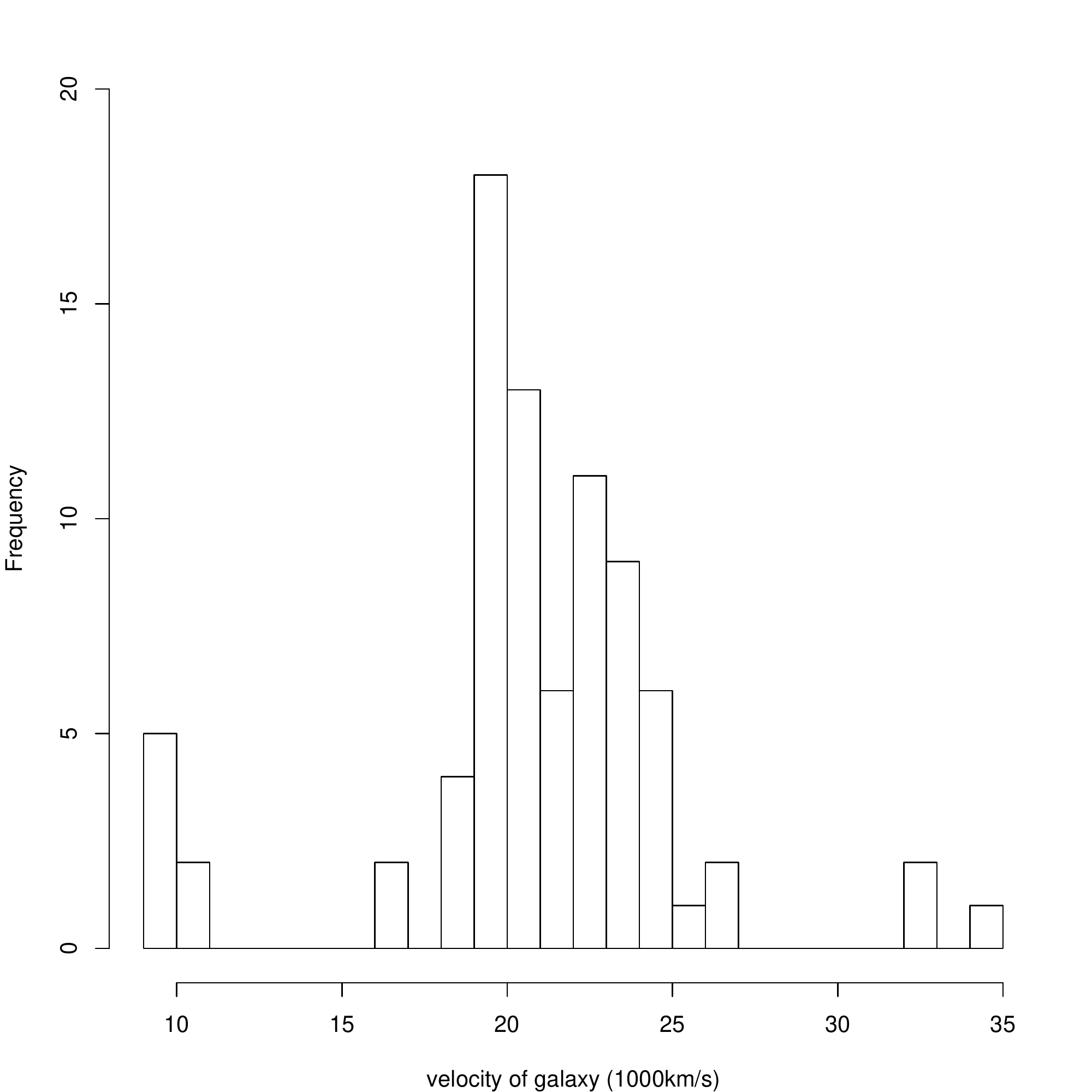}

\caption{Histogram of the velocities of 82 galaxies}
\label{fig:gal}
\end{figure}

The galaxy dataset is commonly used to illustrate mixture modelling techniques \citep[see][for a recent investigation of this dataset in the mixture modelling context]{jasra05}. Briefly, it consists of the velocities of 82 galaxies, but the velocities appear to cluster, suggesting different groups of galaxies that we may wish to identify (see Figure \ref{fig:gal}). If we model these data using a mixture, it is likely that we wish our mixture components to represent the clusters with different mean velocities, hence the PPP of Equation (\ref{eq:ppp-diff}) could be considered in this scenario. We run a variable dimension sampler with the details as above, with normally distributed components assumed and a Poisson$(1)$ prior distribution on the number of components $K$. We compare the results of standard priors \citep[i.e.\ those given in][]{richardson97} with the standard priors plus the PPP. Both with and without the PPP, the values of $K$ with the majority of posterior support are $K=3$ and $K=4$ \citep[but see][for discussion on the posterior of the number of components in a mixture model]{aitkin01}. For the $K=3$ case the posterior means are already well separated, and the PPP has little or no effect on the posterior means. We look in more detail at the $K=4$ case.

\begin{figure}[htp]
\centering

    \includegraphics[scale=0.5] {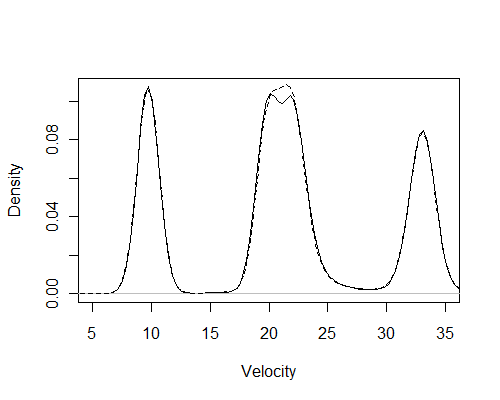}

\caption{Smoothed density of a generic $\mu_k$ for the galaxy data. Without PPP: dashed line; with PPP: solid line.}
\label{fig:gal-muall}
\end{figure}

In order to avoid the label switching issue, we first consider the posterior of a generic $\mu_k$ without relabelling, estimating this by combining into a single vector all samples from the posterior $\mu_k$, for $k=1,2,3,4$, conditional on $K=4$. We can do this since invariance of the posterior under re-parametrisation means we can ignore the labels. The resulting density plot is given in Figure \ref{fig:gal-muall}. The interesting difference to note here is that with the PPP four distinct peaks can be observed in the density, whereas without the PPP the middle two peaks cannot be distinguished.
This does, however, depend on the smoothing parameter used in the non-parametric density estimate.

\begin{figure}[htp]
\centering

    \includegraphics[scale=0.5] {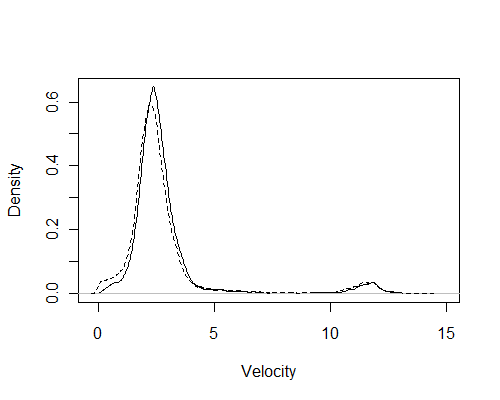}

\caption{Smoothed density of $(\mu_3 - \mu_2)$ for the galaxy data with $K=4$ after an IC is applied. Without PPP: dashed line; with PPP: solid line.}
\label{fig:gal-mudiff}
\end{figure}

To consider this further we mitigate the label switching issue by applying the identifiability constraint $\mu_1 < \mu_2 < \mu_3 < \mu_4$, then look at the posterior density of $(\mu_3 - \mu_2)$. This is given in Figure \ref{fig:gal-mudiff}. We see that applying the PPP causes more separation between the two component means (less mass at small differences).

\section{Discussion} \label{sec:discuss}
In this paper we have introduced the idea of incorporating weak joint information about parameters in a mixture model into the prior specification. In particular we have introduced proximity penalty priors (PPPs) as a method of explicitly declaring an a-priori opinion (or interest) in components that differ on a certain parameter.
The formulation is designed to allow this opinion to be as vague as possible: we avoid making any statement about the magnitude of the difference that should be observed between the components, i.e.\ the method is scale-free.

With the focus of this paper being introduction of the idea, the examples were kept fairly simple. The idea, however, is very general and could be applied in more complex models. For example, in an application such as genetics we may wish to construct a mixture of regressions with many covariates. Suppose there are $p$ covariates and $K$ mixtures, with the coefficient of the \kth{j} covariate in the \kth{k} mixture given by $\beta_{jk}$.  Then we could consider the PPP
$$
p_2(\bm{\gamma}) = \max_j \min_{k \neq l} |\beta_{jk} - \beta_{jl}|,
$$
to reflect a belief that each component should have at least one coefficient that differs from the value in every other component.

Another potential extension is to replace the $L_1$-norm assumed in the PPP with an $L_s$-norm, i.e. considering a generalisiation of, for example, Equation (\ref{eq:simple-mu}), to
\begin{equation*}
p_2(\bm{\gamma}) = |\mu_1 - \mu_2|^s.
\end{equation*}
In this generalised setting, we note that $s=0$ clearly corresponds to an unpenalised prior and $s=1$ reduces to the original Equation (\ref{eq:simple-mu}). Also, setting $s=-1$ encodes a PPP like Equation (\ref{eq:ppp-same}). This generalisation then raises the question of how should $s$ be chosen? We suggest $s=1$ is a very natural choice, since this means the penalty is being applied on the original scale of the data. We have, however, looked at the sensitivity to the choice of $s$. For the example considered in Section \ref{sec:results}, once $s$ becomes large the posteriors for $\bm{\mu}$ become very flat.

\bibliographystyle{apalike}
\bibliography{E:/Documents/Bib/biblio-full}

\end{document}